\documentclass[preprint]{aastex63} 

\usepackage[capbesideposition=right]{floatrow}
\usepackage{wrapfig}
\usepackage[right = 0.5in, bottom = 1in]{geometry}

\newcommand\aastex{AAS\TeX}

\newcommand{\Rs}{$ R_{\odot} $}
\def\ion[#1 #2]{#1\,{\sc #2}}

\submitjournal{ApJ}
\shorttitle{\aastex\ F-Corona Color and Brightness }
\shortauthors{Boe et al.}

\begin{document}

\title{The Color and Brightness of the F-Corona Inferred from the 2019 July 2 Total Solar Eclipse}

\author{Benjamin Boe}
\affil{ Institute for Astronomy, University of Hawaii, Honolulu, HI 96822, USA}

\author{Shadia Habbal}
\affil{ Institute for Astronomy, University of Hawaii, Honolulu, HI 96822, USA}

\author{Cooper Downs}
\affil{Predictive Science Inc., San Diego, CA, 92121, USA}

\author{Miloslav Druckm\"uller}
\affil{Faculty of Mechanical Engineering, Brno University of Technology, Technicka 2, 616 69 Brno, Czech Republic}

\correspondingauthor{Benjamin Boe}
\email{bboe@hawaii.edu}
\begin{abstract}
Total solar eclipses (TSEs) provide a unique opportunity to quantify the properties of the K-corona (electrons), F-corona (dust) and E-corona (ions) continuously from the solar surface out to a few solar radii. We apply a novel inversion method to separate emission from the K- and F-corona continua using unpolarized total brightness (tB) observations from five 0.5 nm bandpasses acquired during the 2019 July 2 TSE between 529.5 nm and 788.4 nm. The wavelength dependence relative to the photosphere (i.e., color) of the F-corona itself is used to infer the tB of the K- and F-corona for each line-of-sight. We compare our K-corona emission results with the Mauna Loa Solar Observatory (MLSO) K-Cor polarized brightness (pB) observations from the day of the eclipse, and the forward modeled K-corona intensity from the Predictive Science Inc. (PSI) Magnetohydrodynamic (MHD) model prediction. Our results are generally consistent with previous work and match both the MLSO data and PSI-MHD predictions quite well, supporting the validity of our approach and of the PSI-MHD model. However, we find that the tB of the F-corona is higher than expected in the low corona, perhaps indicating that the F-corona is slightly polarized -- challenging the common assumption that the F-corona is entirely unpolarized.
\end{abstract} 

\section{Introduction} 
\label{intro}

Coronal emission in the visible wavelength range consists of broadband `continuum' emission, overlaid with a number of discrete spectral lines. Line emission is useful for studying various physical processes in the corona (e.g., \citealt{Habbal2011, Habbal2013, Ding2017, Boe2018, Boe2020a, DelZanna2019}), as it arises from the collisional and radiative excitation of various atomic and ionic species. Line emission is rather faint compared to the total coronal brightness however, and occurs only at well-defined wavelengths, making it straightforward to separate it from the continuum brightness.
\par

Continuum emission in the corona at visible wavelengths originates from two sources, each of which are physically interesting for different reasons. First, there is Thompson scattering of photospheric light by free electrons throughout the corona (e.g., \citealt{vandeHulst1950}), known as the K-corona. Second, there is dust throughout the solar system which scatters and/or diffracts photospheric light (e.g., \citealt{Koutchmy1985}), called the F-corona. Both the K- and F-corona originate via the redirection of photospheric light, and so have a spectrum roughly proportional to the solar photosphere. The wavelength overlap and similar behavior of the K- and F-corona make them somewhat more challenging to isolate from each other, a task which is essential to investigate their physical characteristics. 

\par
The most common method for differentiating K- and F-corona emission has been to compare polarized brightness ($pB$) and total brightness ($tB$) observations (e.g., \citealt{Ney1961, Morgan2007}), since the K-corona is substantially polarized as a result of the scattering geometries \citep{Minnaert1930, Baumbach1938}. The K-corona is not completely polarized however, so isolating the F-corona brightness requires an inversion method to infer the $tB$ of the K-corona with a $pB$ observation and a model of the polarization fraction (see \citealt{Lamy2020} for a detailed discussion). Another, more historical, method to isolate the K- and F-corona emission is by observing the relative depth of Fraunhofer lines in the corona. Fraunhofer lines are the photospheric absorption lines present in the solar spectrum, which are retained in the dust scattering (e.g., \citealt{vandeHulst1950}). The K-corona does not have these lines, as the high kinetic temperature of electrons ($T_e \sim 10^6$ K) in the corona smooths out the scattered solar spectrum in the K-corona. In fact, the names K- and F-corona refer to exactly this effect (in German) -- where K is `Kontinuierliche' refers to its smooth continuous spectrum, while F is `Fraunhofer' for the presence of these photospheric lines. However, some broad Fraunhofer lines such as \ion[Ca II] H and K do leave small temperature dependent variations in the K-corona spectrum \citep{Cram1976} that have been used to infer the coronal electron temperature \citep{Reginald2003, Reginald2009}.

\par
The spatially dependent scattering of the K-corona -- depending both on the plane-of-sky (POS) distance from the Sun, or elongation angle (often referred to as the `impact parameter'), and the location of structures along the line-of-sight (LOS) -- enables the characterization of 3-dimensional (3D) structures like streamers (e.g., \citealt{Kramar2014}) and Coronal Mass Ejections (CMEs; e.g., \citealt{Moran2004, Dere2005, Howard2009}), especially if the event is observed from multiple viewing angles around the Sun (e.g., \citealt{Colaninno2009, Moran2010, DeForest2017}). Such $pB$ inversion methods have been extensively used to infer the electron density in the corona (e.g., \citealt{Allen1947, Guhathakurta1996, Skomorovsky2011, Morgan2020}). Techniques have been developed to infer the electron density from total brightness ($tB$) observations as well \citep{Hayes2001}, though such methods require an assumed F-corona contribution to the continuum. Fine-scale density structures observed in $tB$ emission of the K-corona have also been used to trace the coronal magnetic field topology \citep{Boe2020b}.

\par
While observing K-corona emission is propitious for coronal research, F-corona emission is typically more relevant to solar system science as a constraint on interplanetary dust (e.g., \citealt{Leinert1990}). Indeed, the F-corona is often referred to simply as the component of Zodiacal light that has the smallest elongation angles from Sun center (e.g., \citealt{Kimura1998}). Unlike the K-corona, which originates from the scattering of free electrons close to the Sun, the F-corona is originating from dust along the entire LOS from the Sun to 1 AU and beyond (e.g., \citealt{vandeHulst1947, Mann1998}). This large LOS depth of the F-corona (i.e., Zodiacal light) likely explains why the F-corona has been found to remain almost constant throughout the solar cycle \citep{Morgan2007}. There remains some uncertainty as to the behavior of dust particles near the Sun, and it is expected that the dust should vaporize in the vicinity of the Sun within a distance range referred to as the `dust-free zone' (e.g., \citealt{Mukai1974,Lamy1974}). The size and dynamics of this dust-free zone have not been observationally characterized, but there are early observations from Parker Solar Probe that perhaps support its existence \citep{Howard2019}.

\par

Another important difference between K- and F-corona emission, is color. Throughout this article, we will use `color' to refer to the spectral intensity of the K- and F-corona with respect to the photospheric spectrum at different wavelengths (for each LOS), rather than the absolute intensity. That is to say, we are interested in wavelength dependent scattering efficiency as a way to probe the spatially distributed properties of the dust and electrons (see Section \ref{Eclipse} for more discussion). The K-corona will have a roughly neutral color relative to the photospheric spectrum, as Thompson scattering itself has no wavelength dependence -- though there is a small K-corona color effect resulting from the color dependence of limb-darkening (see \citealt{Qumerais2002, Howard2009}, Section \ref{LimbDark}). Conversely, the F-corona is substantially reddened due to dust diffraction \citep{vandeHulst1947}, and will have a somewhat different color depending on the mineralogy of the scattering particles \citep{Roeser1978}. 

\par
 
In this work, we introduce a novel inversion technique to isolate the K- and F-corona brightness via a color analysis. We used five unique narrowband observations of the coronal continuum between 529.5 and 788.4 nm from the 2019 July 2 Total Solar Eclipse (see Section \ref{Eclipse}), combined with the corresponding Mauna Loa Solar Observatory (MLSO) COMSO K-Cor polarized coronagraph data (see Section \ref{Kcor}). In Section \ref{method} we describe this novel inversion technique, followed by a discussion of our inferred color of the F-corona (see Section \ref{FcorColor}), brightness of the K- (see Section \ref{KcorBright}) and F-corona (\ref{FcorBright}), and compare with the Predictive Science Inc. (PSI) Magnetohydrodynamic (MHD) model of the eclipse corona (see Section \ref{PSI}), as well as earlier published work (see Section \ref{results}). Concluding remarks are given in Section \ref{conc}.

\section{Data}
\subsection{Eclipse Observations}
\label{Eclipse}
\begin{figure*}[t!]
\centering
\epsscale{0.7}
\includegraphics[width = 5.5in]{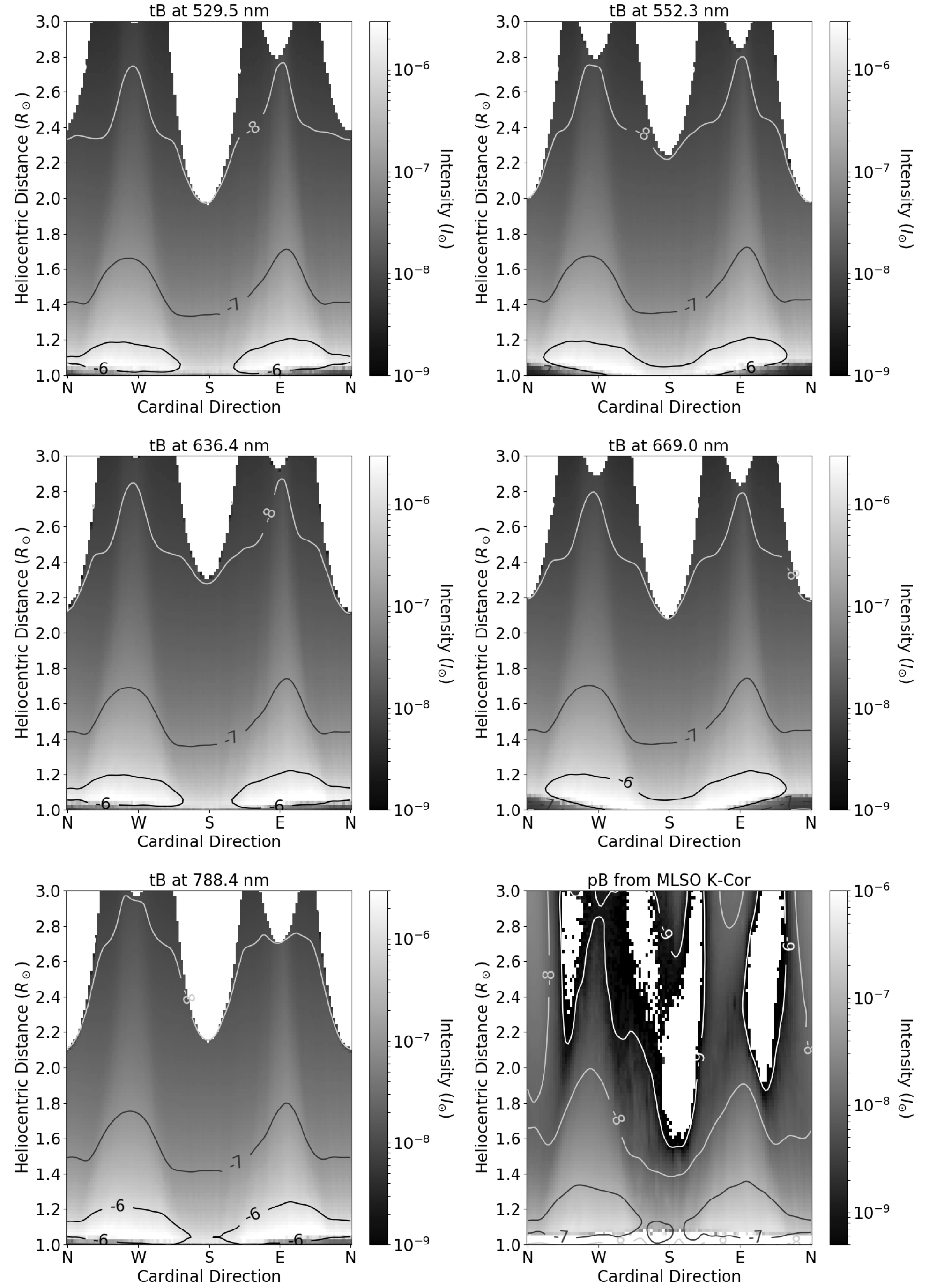}
\caption{Total brightness ($tB$) of the continuum corona at each of the five narrowband wavelengths (see Section \ref{Eclipse}), arranged in increasing wavelength from left to right, top to bottom. The bottom right panel shows the observed polarized brightness ($pB$) of the corona from the MLSO COSMO K-Cor (see Section \ref{Kcor}). Contours are shown for each integer of brightness in log space. The eclipse data are calibrated using the procedure described in Section \ref{method}, and is given in units of solar disk center intensity (for each wavelength). Metadata about the eclipse observations are included in Table \ref{tableEclipse}.}
\label{figEclipse}
\end{figure*}

The eclipse data used in this work were acquired during the 2019 July 2 Total Solar Eclipse in Chile and Argentina. There were three observing sites stationed along the path of totality, to increase the likelihood of clear skies. Specifically, observations were made in Chile at Cerro Tololo (CTIO), and Mammalluca, as well as in Rodeo in Argentina. All the eclipse data used in this work are shown in Figure \ref{figEclipse}, with observing metadata included in Table \ref{tableEclipse}. The data as shown in this figure are transformed into a cartesian representation of polar coordinates by binning the original data into wedges with radial steps of 0.02 \Rs \ and a width of $3^{\circ}$ in latitude. The photometric imaging data are averaged inside each of these bins to increase the signal-to-noise ratio (SNR). This procedure will cause more pixels to be co-added at larger heliocentric distances, which will help to compensate for the decreasing SNR with distance (due to the decreasing coronal brightness).  

\par

\par
Each telescope is outfitted with a 0.5 nm bandpass Fabry-P\'erot filter to record the coronal emission at a precise wavelength. To isolate emission from the ions alone, each ionic line that is observed has an additional imaging system with a bandpass shifted to the nearby continuum. In making these continuum observations, we have serendipitously recorded the relative continuum emission in the corona at well defined wavelengths across the visible spectrum, which is the focus of this work. We will not consider the line emission observations here. A detailed description of the design and operation of these narrowband Fabry-P\'erot telescopes is given in \cite{Boe2020a}. The exact same procedure is applied to the data in this work. We use some of the same telescopes (i.e., \ion[Fe xi], \ion[Fe xiv] and \ion[Ar x], see Table \ref{tableEclipse}) that were described in \cite{Boe2020a}, as well as a few additional ones (i.e., \ion[Fe x] and \ion[Ni xv]). The image alignment and stacking procedure is described in \cite{Druckmuller2009}. 
\par
The sky brightness is removed by measuring and subtracting the intensity observed at the center of the Moon during the eclipse. There will be a slight brightness from the Moon due to Earthshine, which will vary depending on the geography of the Moon facing side of the Earth (e.g., oceans are darker than land or snow). From the calculations of \cite{Agrawal2016}, we take the Earthshine to be $2.5 \times 10^{-10}  \pm 1.5 \times 10^{-10}  I_\odot$ for all wavelengths (it will be a continuum source like the corona itself). This intensity is rather small compared to the lower corona, so this estimate will only have a noticeable impact on the results at high elongations ($\sim$ 2.5--3 \Rs).

\begin{deluxetable}{cccccc}
\tablecaption{Observing metadata for the 2019 July 2 Total Solar Eclipse observations used in this work.}

 \label{tableEclipse}
\tablewidth{5in}
\tabletypesize{\small}
\tablehead{
\colhead{$\lambda (nm)$} & \colhead{Aperture (cm)} & \colhead{Line}  & \colhead{Observing Site} & \colhead{Length of Totality} & \colhead{Observers}} 
\startdata
529.5  & 7 & \ion[Fe xiv] & Rodeo, Argentina & 2m 14s & Judd Johnson and Pavel \v Starha\\
\hline
552.3  &  7 & \ion[Ar x] & Mammalluca, Chile  & 2m 21s & Martina Arndt, Rydia Hayes, \\ & & & & & Sarah Auriemma and Benedikt Justen \\
\hline
636.4  &  7 & \ion[Fe x] & Rodeo, Argentina & 2m 14s & Judd Johnson and Pavel \v Starha\\
\hline
669.0 & 5  & \ion[Ni xv] & Cerro Tololo, Chile & 2m 3s & Petr \v Starha and Jana Hod\'erova \\
 \hline
788.4  & 7 &  \ion[Fe xi] & Rodeo, Argentina & 2m 14s & Judd Johnson and Pavel \v Starha\\
\hline
\enddata
\end{deluxetable}

To photometrically calibrate the eclipse data we use the MLSO COSMO K-Cor data (see Section \ref{Kcor} and Fig. \ref{figEclipse}) after they have been corrected by the polarized brightness fraction ($pB$/$tB$) from the PSI-MHD model (see Section \ref{PSI}) and corrected for limb-darkening effects (see Section \ref{LimbDark}). The K-Cor data are calibrated to be in units of the solar disk center intensity already, so they can be used as a calibration source for the eclipse data -- though the eclipse observations are able to probe significantly farther into the corona (out to $> 2.5$ \Rs) than the MLSO coronagraph ($< 1.5$ \Rs). To do this cross calibration, we take the intensity in a $10^\circ$ wedge from 1.125 to 1.15 \Rs \ in the Eastern streamer, since it had the lowest F/K ratio, and normalize all the eclipse data to the K-Cor data in this wedge.  Implicit in this assumption is that the F-corona intensity is negligible within this wedge, which may not be the case. In Section \ref{derivation}, we present a method for isolating the F-corona in the $tB$ eclipse observations, then iterate the procedure to account for the F/K calibration offset in Section \ref{iteration}. 
\par
It is important to note that this calibration process is performed in units of the solar disk center intensity at the given wavelength, which normalizes any dependence of the solar disk spectrum on the brightness observations presented here. In this way, we are working with the relative scattering of the incoming solar radiation, which provides information about the physical properties of the scattering particles (i.e., dust and electrons) that we are interested in here. The absolute intensity of the corona will depend on the incoming solar photospheric spectrum as well, but the absolute intensity itself is not directly relevant to this work. 
\par
For wavelengths in the vicinity of rather large Fraunhofer lines, there will be a temperature dependent effect where the K-corona spectrum can change somewhat due to a variable broadening of the lines. In fact, this effect has been previously used to infer the electron temperature in the low corona with spectroscopic eclipse observations \citep{Reginald2003, Reginald2009}. However, this effect will only be noticeable for small elongation angles ($< 1.3$ \Rs) and in the vicinity of very deep Fraunhofer lines such as \ion[Ca II] H and K (396.9 and 393.3 nm respectively; \citealt{Cram1976}), which is not the case for the bandpasses used in this work. Moreover, our calibration is relative to the disk center intensity integrated over the bandpass (via the MLSO K-Cor cross-calibration), so any effect from sufficiently small Fraunhofer lines on the integrated intensity is already accounted for. 
\par
This method of cross-calibration also accounts for the wavelength dependent absorption by the Earth's atmosphere in the observations, as the observed intensity within the calibration wedge itself is convolved with the atmospheric attenuation. The atmospheric effects will be angle dependent across the sky, but we can assume a constant absorption fraction across the eclipse images, given that the field-of-view is within the small-angle approximation regime.

\subsection{MLSO COSMO K-COR}
\label{Kcor}
In addition to the observed total brightness of the corona at the various wavelengths during the 2019 TSE (as described in Section \ref{Eclipse}), we used the ground-based coronagraph observations from the Mauna Loa Solar Observatory's (MLSO) COronal Solar Magnetism Observatory (COSMO) K-coronagraph (K-Cor) \footnote{K-Cor DOI: 10.5065/D69G5JV8; \url{https://mlso.hao.ucar.edu/mlso_data_calendar.php}}. The COSMO K-Cor instrument observes the polarized brightness ($pB$) of the corona on a daily basis in the lower regions of the corona starting above $\sim1.1$ \Rs \ with a wavelength bandpass between about 720 and 750 nm. The design and characterization of the polarimeters for K-Cor are described in \cite{Hou2013}. The K-Cor data are calibrated to solar disk center intensity units, and so they are a useful source for cross-calibrating the eclipse data (as done in Section \ref{method}). 
\par
To improve the quality of the K-Cor data, we averaged all available images for the entire day of 2019 July 2, as shown in Figure \ref{figEclipse}. Stacking the K-Cor data helped to reduce any sky brightness effects, and to increase the signal of the corona. Still, the K-Cor data clearly has non-physical features present above about 1.3 \Rs \ in some regions of the data, probably due to imperfect subtraction of the Earth's atmospheric brightness (see Section \ref{KcorBright}). This approach will smooth any changes in the corona over this period. We can safely assume a static corona over this time period however, as we do not expect any large changes over a few hours from the solar rotation and there were no CMEs observed on this day.

\subsection{PSI-MHD Model}
\label{PSI}
In addition to the observational data described earlier in this section, we also make use of a state-of-the-art high-resolution Magnetohydrodynamic (MHD) model of the global corona on 2019 July 2, posted by the PSI group one week before totality\footnote{\url{www.predsci.com/eclipse2019}}. This simulation used a similar setup to the 2017 August 21 eclipse prediction, which is described in \cite{Mikic2018}. This includes a Wave-Turbulence-Driven (WTD) approach to specify coronal heating and a novel energization technique to add field aligned currents (shear/twist) over large-scale polarity inversion lines. Based on a quantitative comparison of the 2017 prediction to coronagraph observations \citep{Lamy2019}, small updates were made to the 2019 parameterization of the WTD model to increase the overall electron density of the model. This included a slight increase in the WTD Poynting flux at the inner boundary (increased heating), and a switch to using hybrid abundances in the radiative loss calculation \citep{Schmelz2012, Landi2012}. 
\par
To model the corona at a given time, a full-sun map of the radial component of the photospheric magnetic field at the inner boundary must be specified. Because of the lead-time required for publishing the prediction, this simulation used a splice of synoptic map data from SDO/HMI \citep{Scherrer2012} based on what was available about 15 days prior to the eclipse. This map combined data for Carrington rotation 2217 with near-real-time data from a part of Carrington rotation 2218. To capture plume-like density structures at the poles, the polar caps were also filled with a random flux distribution whose net value matches observations of the net flux at high latitudes (also similar to \citealt{Mikic2018}). All told, the measurements span approximately 2019 May 22 to June 17, which places the oldest data near the west limb during totality. 
\par

\begin{figure*}[t!]
\centering
\epsscale{0.7}
\includegraphics[width = 5.5in]{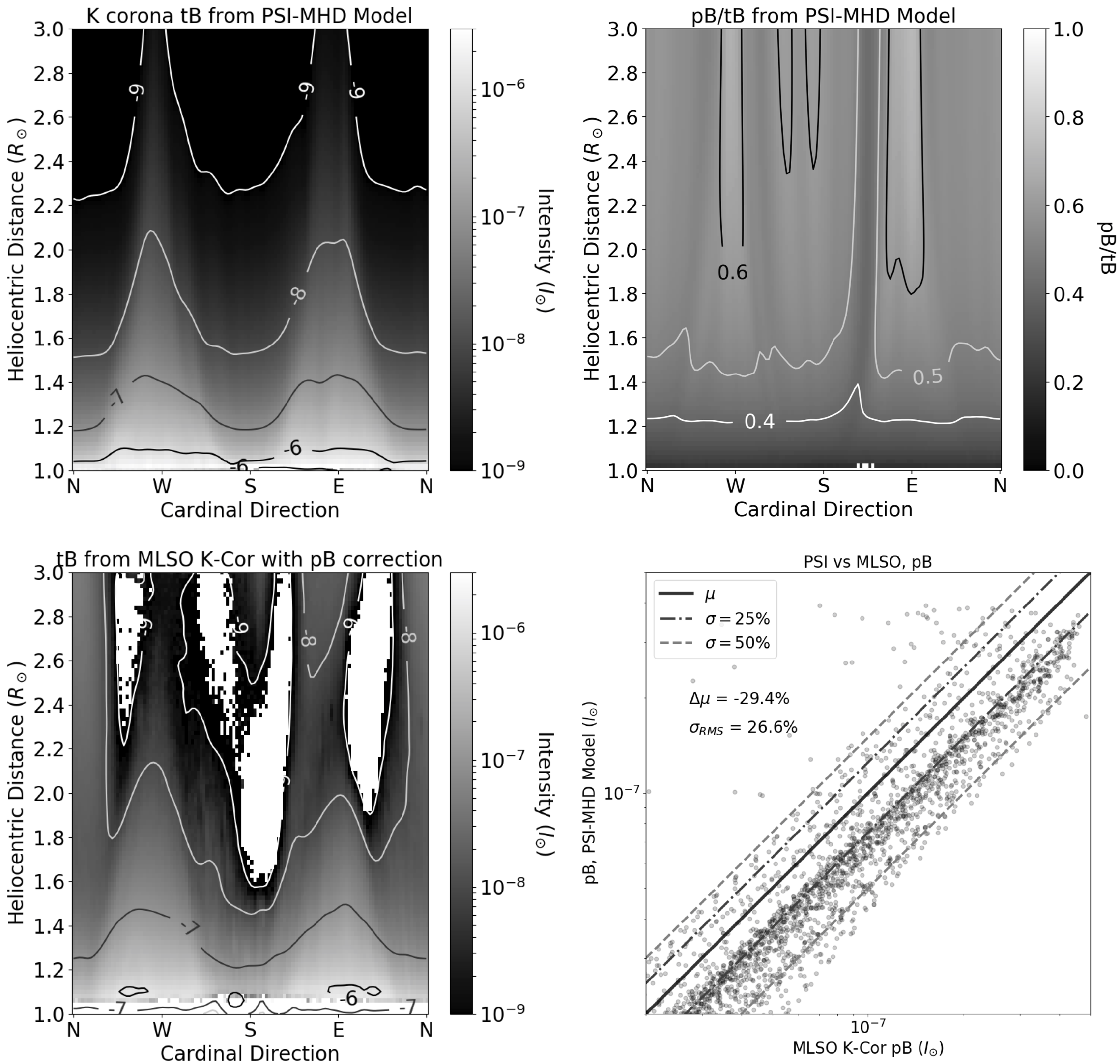}
\caption{Top left: Predicted Total Brightness ($tB$) of the K-corona from the PSI-MHD model at 529.5 nm (see Section \ref{PSI}). Top right: $pB$/$tB$ ratio at 529.5 nm from the PSI-MHD model. Bottom left: Inferred $tB$ by correcting the K-Cor $pB$ data (see Fig \ref{figEclipse}) with the model $pB$/$tB$ ratio and limb darkening effects (see Section \ref{LimbDark}). Bottom Right: Comparison between the K-Cor data and model $pB$.}
\label{figPSI}
\end{figure*}

Here we will use the forward modeled total and polarized line-of-sight (LOS) integrated brightness for the K-corona from the model. It is important to remark that the PSI-MHD model does not contain any dust, and so is a prediction of the K-corona intensity alone, without any contribution from the F-corona. The $tB$ and $pB$ of the K-corona (for 529.5 nm) according to the model are shown in Figure \ref{figPSI}, with the same polar coordinate transformation as applied to the eclipse data in Figure \ref{figEclipse} (see Section \ref{Eclipse}). A direct comparison between the predicted $pB$ and the $pB$ observed by MLSO COSMO K-Cor (see Section \ref{Kcor}) is shown in the lower right panel of the figure. The model predicted and observed $pB$ show roughly the same slope in brightness, but with the MLSO data being about $29.4\%$ brighter on average than the model -- which is just barely statistically significant given the scatter being $26.6 \%$ around the mean.

\par

The high resolution MHD model from PSI is useful for comparison with photometric observations of the corona (see Sections \ref{results} and \ref{conc}), which perhaps can provide insight on the limitations and constraints of the model. The model can also be used as a reasonable estimate of the polarization fraction for the K-corona scattered light. While there may be inaccuracies in the model or its assumptions (which we want to test ultimately), the polarization fraction is somewhat insensitive to the absolute values of density in the model, and is much more influenced by the geometry of the Sun and corona as extended 3D sources. We can thus use the predicted $pB/tB$ from the PSI model combined with the observed $pB$ from MLSO's K-Cor to infer the expected $tB$ of the K-corona based on the observations. This correction will create a K-Cor driven inference of $tB$ that is directly comparable to the eclipse inference. The $pB/tB$ ratio from the PSI model as well as the corrected $tB$ inference based on the K-Cor data (after accounting for limb darkening effects, see Section \ref{LimbDark}) are shown in the top right and bottom left panels of Figure \ref{figPSI} respectively.

\newpage
\section{K- and F-corona Inversion}
\label{method}
The continuum data from the solar eclipse (see Section \ref{Eclipse}) contain emission from the K- and F-corona combined at very specific wavelengths. Hence, the relative emission observed at each bandpass will depend on the spectrum of these two continuum sources. In the inversion method presented here, we used the color spectrum (relative to the photosphere) of the F-corona to isolate the K- and F-corona contributions to the total continuum emission. The K-corona itself also has a slight color, due to the wavelength dependence of limb-darkening, and so our inversion method accounts for the color of both the K- and F-corona. The determination of the K-corona color is done in Section \ref{LimbDark}, which is followed by the derivation of the analytical inversion approach in Section \ref{derivation}. This method is then iterated (in Section \ref{iteration}) to find the best fit solution of the F-corona color and relative intensity of the K- and F-corona.

\subsection{Limb-Darkening}
\label{LimbDark}
The surface intensity of the solar photosphere is not uniform in intensity at a given wavelength, but rather depends on the physical depth at which the solar disk is optically thick (which varies across the disk) -- combined with the temperature distribution of the solar atmosphere. The photospheric radiation at the limb of the Sun is substantially fainter and redder (for visible wavelengths) than the center of the solar disk, and so this limb-darkening effect will impact the relative irradiance entering the corona in an angle dependent manner. Thompson scattering itself has no wavelength dependence, but the angle and wavelength dependence of limb-darkening coupled with an extended 3D corona will result in a variable intensity of the K-corona at different wavelengths and heliocentric distances (e.g., \citealt{Qumerais2002}). The traditional prescription for limb-darkening (see \citealt{Minnaert1930, Howard2009}) is given by the following:
\begin{eqnarray}
\label{LimbDarkEqn}
I(\psi) = I_0 (1 - u + u \cos \psi),
\end{eqnarray}
where $u$ is the limb-darkening coefficient, and $\psi = \sin^{-1} R/R_\odot$ with $R/R_\odot$ being the projected distance from disk center as viewed from the Earth (i.e., small-angle approximation). We refer the reader to \cite{Howard2009} for a full discussion on the scattering physics in the corona and the impact of limb-darkening. 
\par
\begin{figure*}[t!]
\centering
\epsscale{0.7}
\includegraphics[width = 6.5in]{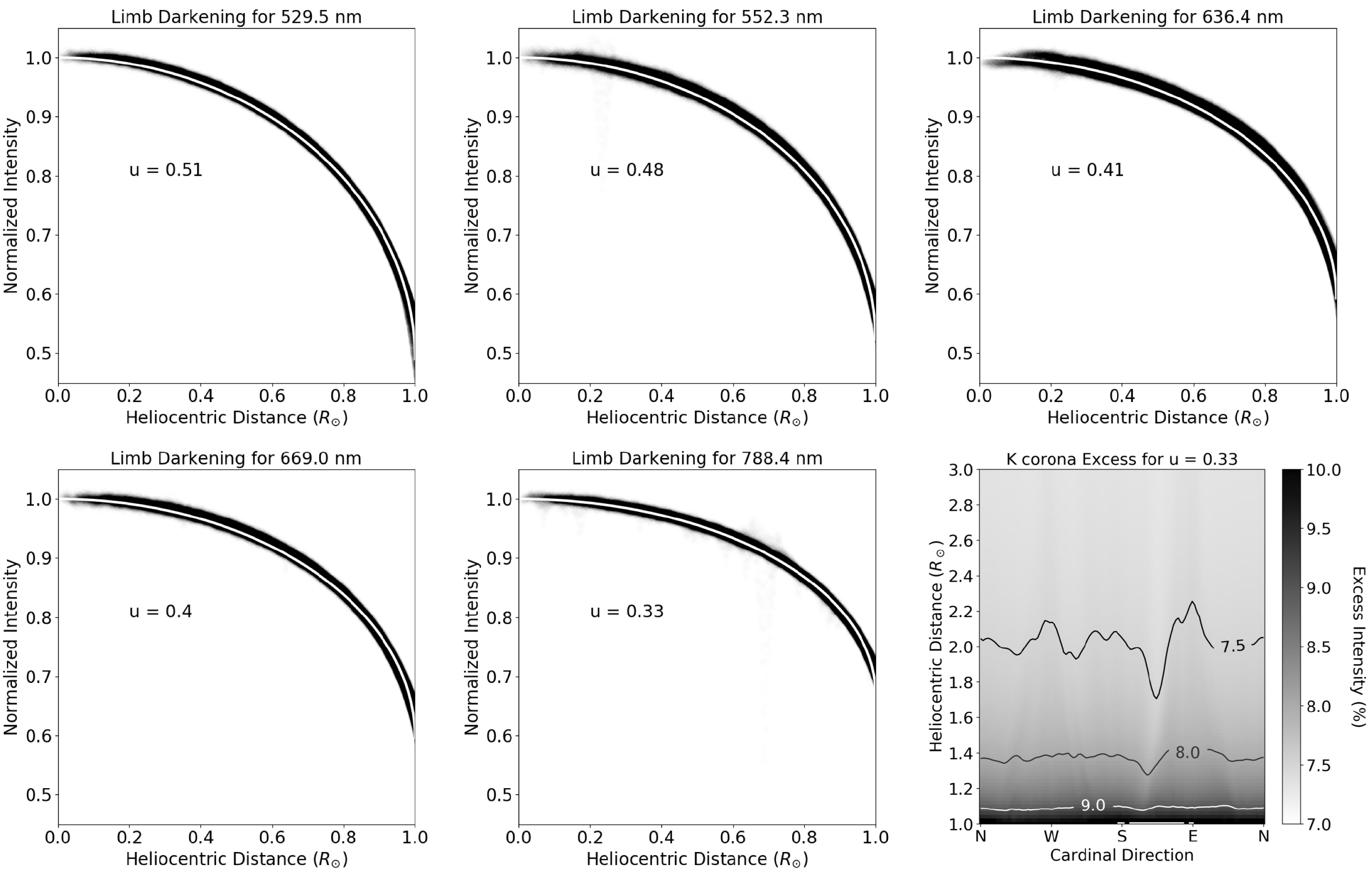}
\caption{Limb-darkening data for each of the eclipse bandpasses, arranged in increasing wavelength. Each panel shows the observed disk intensity with black data-points (normalized) as a function of the projected heliocentric distance. The overlaid white line is the best fit to the data using Equation \ref{LimbDarkEqn}, with the best fit limb-darkening coefficient ($u$) noted in each panel. The bottom right panel shows the percentage of excess emission expected for 788.4 nm compared to 529.5 nm from the PSI-MHD model, after normalizing for the solar disk center intensity at each wavelength.}
\label{figLimbDark}
\end{figure*}

To quantify the limb-darkening coefficient for each of the eclipse observations, we fit equation \ref{LimbDarkEqn} to observations of the solar disk performed with each of the narrowband telescopes. These solar disk observations are normally used for relative calibrations between the on- and off-band telescopes for the purposes of measuring line emission (see Section \ref{Eclipse}), which we are not considering here. In Figure \ref{figLimbDark}, we show the observed intensity of the solar disk for each wavelength across the limb, and find the best fit $u$ for each. We find that the limb-darkening coefficient is decreasing with increasing wavelength, resulting in a redder limb compared to disk center (as expected). In fact, we find that the $u$ coefficients had a simple negative linear dependence with wavelength, given by: 
\begin{eqnarray}
\label{EqnUfit}
u(\lambda) = -6.8 \times 10^{-4} \left( \frac{\lambda}{  \mathrm{nm}} \right) + 0.86,
\end{eqnarray}
with a root-mean-square (RMS) error of $\sigma_{RMS} < 2.5\%$ on the fit.

\par
Forward modeling $tB$ from the PSI-MHD model requires a limb-darkening assumption (i.e., $u$), so we computed the LOS integrated K-corona intensity with the best fit $u$ of each bandpass. We then determined the excess emission that is expected at each bandpass relative to the lowest wavelength channel (529.5 nm), after normalizing for the solar disk center intensity at each wavelength. The color changes between the various channels are rather small, with the larger changes occurring for larger wavelength separations. The bottom right panel in Figure \ref{figLimbDark} shows the change in intensity from the lowest to highest wavelength used in this study, which increases up to a maximum of 10\% in intensity for the very low corona.
\par
In the inversion method described in Section \ref{derivation}, we will use the PSI-MHD model prediction for the K-corona color of both $pB$ and $tB$, using the best fit coefficients from solar disk data in Figure \ref{figLimbDark}. All figures in this work which show PSI-MHD model data are given for the limb-darkening coefficient at 529.5 nm. Similarly, the MLSO K-Cor $pB$ data (see Figure \ref{figEclipse} and Section \ref{Kcor}), are presented here after a correction to $pB$ at 529.5 nm. The correction is performed assuming $u = 0.36$ for 735 nm (i.e., middle of K-cor bandpass) using the linear fit from Equation \ref{EqnUfit}. The K-cor data is then converted to $tB$ (as shown in Fig. \ref{figPSI}) using the $pB/tB$ ratio at 529.5 nm. The eclipse data are somewhat more complicated to correct for, as they are convolved with the F-corona intensity (which is wavelength dependent). A more careful approach is required to account for the K-corona color effect in the eclipse data, which is incorporated into the method described in Sections \ref{derivation} and \ref{iteration}.

\subsection{Method Derivation}
\label{derivation}
\par
In this section, we derive an analytical inversion method for isolating the K- and F-corona emission components for a set of total brightness observations at distinct wavelengths. First, we used the K-corona $tB$ provided by the MLSO K-Cor data (see Section \ref{Kcor}) with the $pB$/$tB$ (see Section \ref{PSI}) and limb-darkening corrections (see Section \ref{LimbDark}) as an initial estimate of the K-corona intensity, $K = I_{K} (r, \theta)$, with heliocentric distance $r$ and position angle $\theta$ of an arbitrary line-of-sight (LOS) at a reference wavelength, and in units of the solar disk center intensity. The initial cross-calibration procedure for the data was described in Section \ref{Eclipse}, which will be iterated later in Section \ref{iteration}
 
\par
The intensity of an arbitrary narrowband continuum eclipse image (see Section \ref{Eclipse}) can then be given by:
\begin{eqnarray}
\label{ArbIntensity}
I_i (r, \theta) && = K d_i + F  c_i 
\end{eqnarray}
where $c_i$ and $d_i$ are the color coefficients of the F- and K-corona respectively at the given wavelength, $i$, and $F = I_{F} (r,\theta)$ at the reference wavelength, similar to $K$ above. The K-corona color term is determined by the limb-darkening analysis done in Section \ref{LimbDark}, and is a function of the LOS, thus $d_i = d_i(r,\theta)$. For the sake of this work, we performed the color analysis of the F-corona with respect to the 529.5 nm channel, with an intensity $I_{529.5}$, that will be used as the arbitrary reference wavelength, so $I_{529.5} = I_{0}$. The color of the F- and K-corona at the reference wavelength is then $c_{0} \equiv 1$ and  $d_{0} \equiv 1$. Further, this derivation uses the assumption that all photometric observations are in units of solar disk center intensity for each arbitrary wavelength, and can be thought of simply as the scattering efficiency of each particle species for each LOS.
\par
Solving for the F corona at the reference wavelength is trivial if one has an estimate of the K corona, where:
\begin{eqnarray}
\label{F_529}
F = I_{0} - K.
\end{eqnarray}

\par
Using equations \ref{ArbIntensity} and \ref{F_529} for the 529.5 nm and another arbitrary eclipse observation produces the following:

\begin{eqnarray}
I_i = K \ d_i + (I_{0} - K) c_i,
\end{eqnarray}
then solving for the color term yields,
\begin{eqnarray}
\label{ColorEqn}
c_i = \frac{I_i - K \ d_i}{I_{0} - K}.
\end{eqnarray}

Equation \ref{ColorEqn} thus gives a method for inferring the relative color of the F-corona by comparing an estimate of the K-corona intensity with two arbitrary narrowband observations. Since this derivation applies to a LOS in each image, the method is repeated separately for each LOS and averaged in each combination of data. That is to say, the procedure is repeated for every eclipse image relative to 529.5 nm to solve for the F-corona color, which is averaged across the entire field-of-view. It is possible that there could be some real color variation in the F corona for different LOS positions, which would imply a variable dust-free zone depending on the grain species (see Section \ref{FcorColor}), but we will assume here that the color term is constant within our field-of-view ($< 3 $ \Rs) for each wavelength. The variation in the inferred color term over the dataset then enables a determination of the uncertainty of the average. 

\par
Once the color terms have been measured for the other eclipse wavelengths (i.e., equation \ref{ColorEqn}), the relative F-corona intensity can be inferred separately using each available bandpass. The ratio between the arbitrary and reference images gives,
\begin{eqnarray}
\frac{I_i}{I_{0}}  = \frac{K d_i + F c_i}{K + F}, 
\end{eqnarray}
which can be rewritten to isolate the relative F-corona intensity term (i.e., $\frac{F}{K + F}$), 

\begin{eqnarray}
\frac{I_i}{I_{0}}  && = \left(1 - \frac{F}{K + F} \right) d_i +   \frac{F c_i}{K + F} \\
\frac{I_i}{I_{0}} && = d_i +  \left(c_i - d_i \right)\frac{F }{K + F}   
\end{eqnarray}
and finally solved for the F-corona intensity alone, after using $I_0 = K + F$, 
\begin{eqnarray}
\frac{F }{K + F} && =  \left( \frac{I_i}{I_{0}} - d_i \right)  \frac{1}{c_i - d_i}  \\
F &&=  \left(I_i - I_{0} d_i \right)  \frac{1}{c_i - d_i}  \label{F_arb}
\end{eqnarray}

Equations \ref{F_529} and \ref{F_arb} offer two different methods to derive the same quantity, namely the relative F-corona intensity at the 529.5 nm channel. This can be done directly with the 529.5 nm channel and the K-Cor data in Equation \ref{F_529}, and again for every other eclipse bandpass using Equation \ref{F_arb}. 

\par
We can then extend this procedure to isolate a new map of the K-corona (at 529.5 nm) separately, starting with a modified version of Equations \ref{ArbIntensity} and \ref{F_arb}:
\begin{eqnarray}
K &&= \frac{I_{i} - F c_i}{d_i}    \\[10pt]
K &&= \frac{I_{i}}{d_i} - \left(\frac{I_i}{d_i} - I_{0} \right) \frac{c_i}{c_i - d_i}  \label{K_derive}
\end{eqnarray}

The technique derived in this section provides the means to isolate the relative F- and K-corona contributions as well as the relative color of the F-corona. So far, we have made the assumption that the F-corona is negligible at the base of the streamer (in the calibration wedge discussed in Section \ref{Eclipse}). To address the possible F-corona offset in this calibration wedge, and the impact it has on our inversion procedure, we use an iterative method that is described in Section \ref{iteration}.

\subsection{Iteration Procedure}
\label{iteration}
\par
The process described in the previous section works for coronal regions where the data are reliable between all data sets. For the eclipse data, we can observe emission typically between about 1.05 up to 3 \Rs. The K-Cor data is not as extended however, as they must contend with the high sky brightness that is present in the absence of a total solar eclipse. Additionally, this procedure initially assumes that there is no F-corona contribution to the calibration wedge (see Section \ref{Eclipse}) in the low region of the eastern streamer, since the F-corona offset is initially difficult to immediately estimate without an additional constraint.

 \par
We thus use an iterative process that will isolate the K- and F-corona intensity. The first iteration uses the MLSO K-Cor data as the initial K-corona intensity, $I_{K} (r,\theta)$, and solves for $K$ and $F$ using the 529.5 nm eclipse data and an arbitrary bandpass. Each other eclipse bandpass repeats this procedure separately, as $I_i$, which then yields a different color term for each channel, $c_i$. Each of the other bandpasses is then used to independently generate an inferred map of K and F at 529.5 nm. After each iteration, we have an estimated color for each bandpass and an average map of the K- and F-corona. The newly inferred map of the K-corona is then used as the input for $I_{K} (r,\theta)$ in the next iterative step. In this way, we can use the MLSO K-Cor data as a calibration source and as a means to create an initial estimate of the color terms and K-corona intensity. 

\par
\begin{figure*}[h!]
\centering
\epsscale{0.7}
\includegraphics[width = 3in]{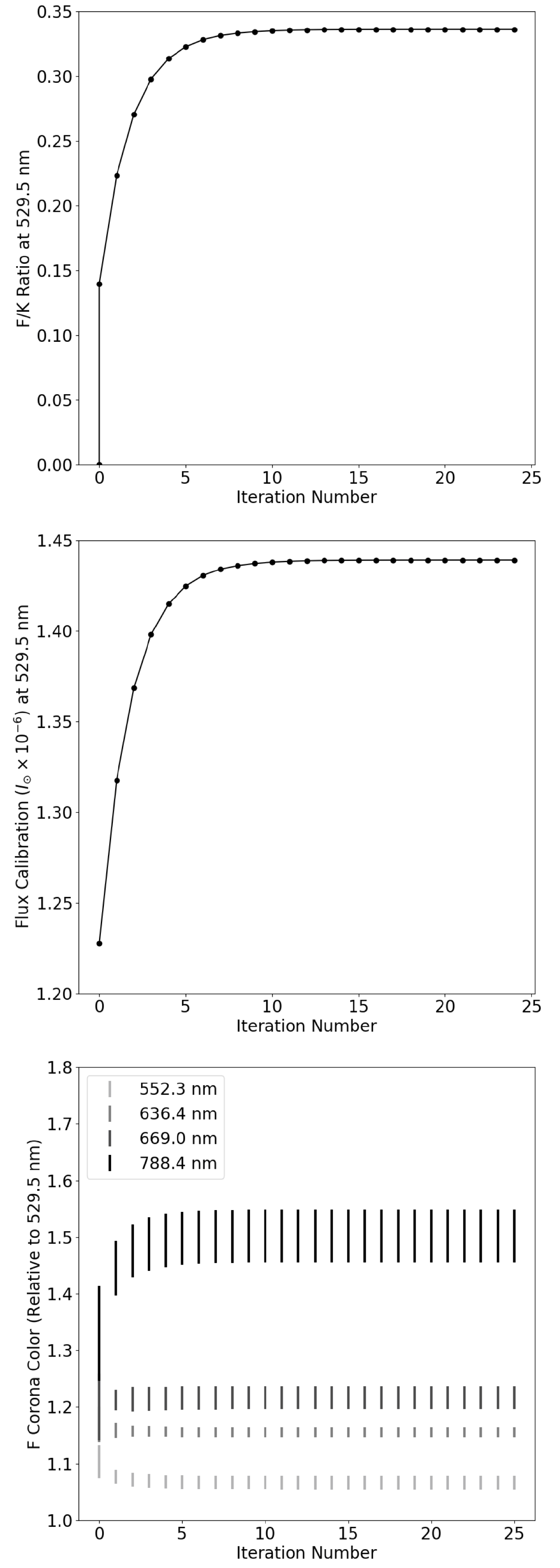}
\caption{Top: F/K ratio for 529.5 nm inferred in the calibration wedge (see Section \ref{Eclipse}) over 25 iterations (see Section \ref{method}). Middle: Flux calibration for 529.5 nm in the calibration wedge after each iteration, other wavelengths have their calibration account for the K- and F-corona colors (see Sections \ref{LimbDark} and \ref{iteration}). Bottom: Corresponding estimate of the F-corona colors for each observed wavelength over the iterations.}
\label{FigIter}
\end{figure*}

Next, we take the inferred F-corona map (from Equation \ref{F_arb}) and fold it twice along the polar and equatorial axes. The result is a latitude averaged, axis-symmetric map of the F-corona. The relative F/K intensity is then measured inside the original calibration wedge, and is used to estimate the F-corona contribution to the brightness inside this calibration wedge (i.e., the flux calibration, see Section \ref{Eclipse}), which in the first iteration had assumed not to have any F-corona intensity. The inferred excess intensity is added to the original data to re-calibrate the images based on the new best guess of the colors and F/K offset in the calibration box. In addition to correcting the F/K offset at each iteration, we correct for the limb-darkening effect on the calibration -- that is, we account for the excess intensity expected in the K-corona at the different wavelength channels given the F/K ratio and K-corona emission in the calibration box. The newly calibrated data combined with the inferred K-corona map are used to repeat the entire procedure of Section \ref{derivation}. After several iterations, the F-corona offset and inferred colors settle on the most probable solution, since the F- and K-corona remain constant under the assumption that the F-corona is axis-symmetric around both the poles and equator. Values for the F/K offset in the calibration wedge, the flux calibration correction, and inferred colors for each iterative step are shown in Figure \ref{FigIter}.

\par

We performed a total of 25 iterations, at which point the solution did not vary within the photometric uncertainty. The value of the F/K ratio changed a lot initially, but quickly converged to a value of $0.336 \pm 0.006$ (the uncertainty is determined by measuring the scatter of the inference for the pixels inside the calibration wedge). Despite the changing F/K offset, the colors at each iteration are rather stable. The scatter (and thus the inferred uncertainty) of the color values drop after a couple of iterations while remaining within about $2 \sigma$ of the original estimate (from the K-Cor data). The colors do spread out slightly from the initial estimate, with the lowest wavelength channel slightly decreasing in the inferred color, and vice versa for the highest wavelength. 

\section{Results}
\label{results}

The novel K- and F-corona inversion technique introduced here (in Section \ref{method}) provides an inference of the F-corona color spectrum (Section \ref{FcorColor}) as well as the spatially resolved total brightness of both the K (Section \ref{KcorBright}) and F (Section \ref{FcorBright}) components of the coronal continuum emission. What follows is a discussion of the results from this inversion technique applied to the 2019 July 2 Total Solar Eclipse data (see Section \ref{Eclipse}), as well as the simultaneous MLSO COSMO K-Cor observations (see Section \ref{Kcor}). These results are then compared with the PSI-MHD model prediction of the eclipse corona (see Section \ref{PSI}) and various historical works. In general, all datasets and the model are reasonably consistent with each other and with previous studies, but there are some interesting deviations that could have implications for the nature of the F-corona.  

\subsection{F-corona Color}
\label{FcorColor}
\begin{figure*}[t!]
\centering
\epsscale{0.7}
\includegraphics[width = 4in]{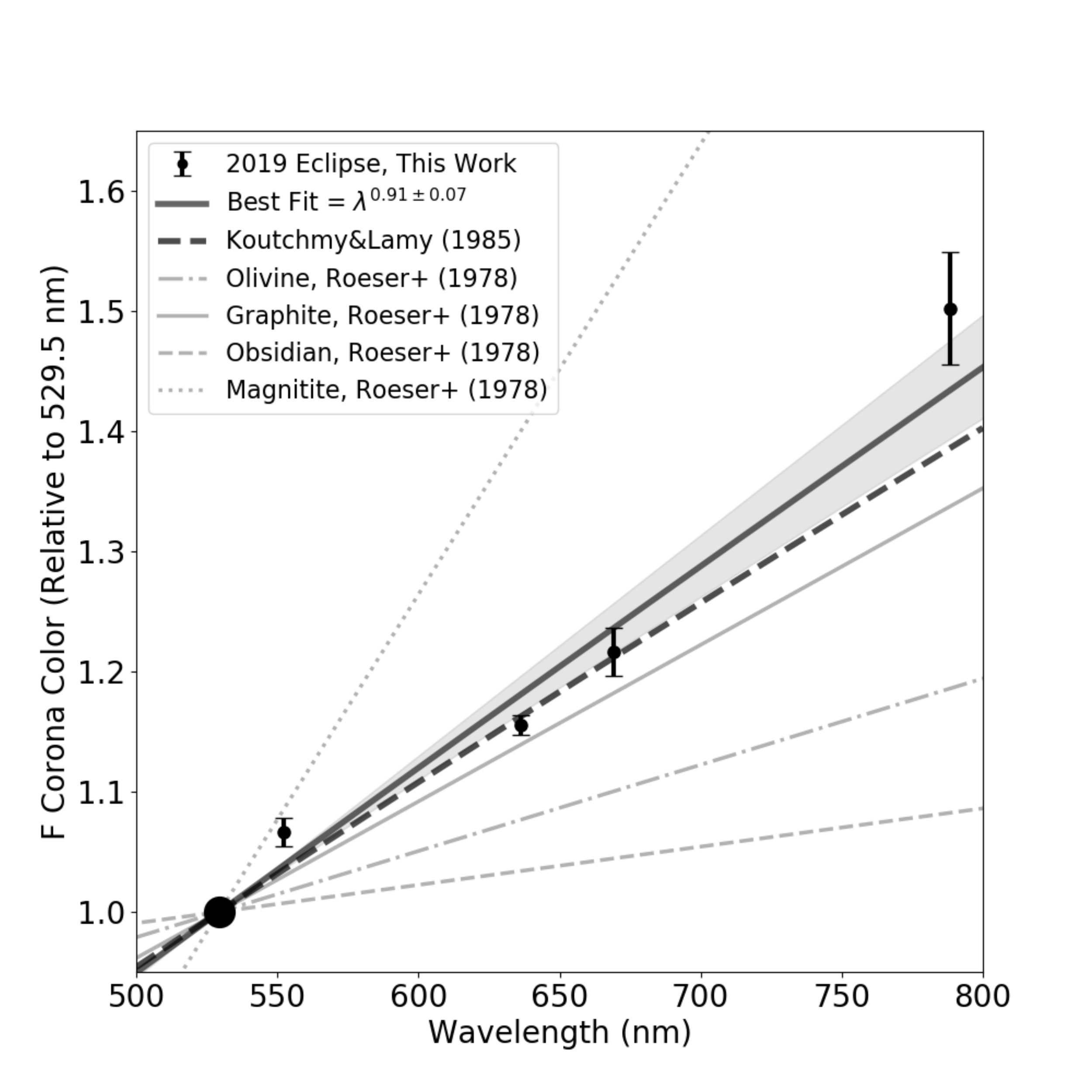}
\caption{Final inference of the F-corona color spectrum using the method outlined in Section \ref{method}. The solid line and the shaded region show the best fit power law and uncertainty of the spectrum based on our analysis. The bold dashed line shows the spectrum presented in \cite{Koutchmy1985}, while the set of thin grey lines show the predicted spectrum for the F-corona given different dust compositions from \cite{Roeser1978}.}
\label{FigColor}
\end{figure*}

The final inferred color spectrum of the F-corona from our inversion method is shown in Figure \ref{FigColor}. A power law fit to the data gives $I_F \propto \lambda^{0.91 \pm 0.07}$, which is roughly consistent with a linear dependence of the color spectrum for visible wavelengths. The fit to our inferred F-corona spectrum is broadly consistent with the slope given by \cite{Koutchmy1985}, which is itself a fit to a few earlier observational studies. Both fits are somewhat close to the predicted spectrum if the F-corona is dominated by graphite grains \citep{Roeser1978}. Graphite grains are expected to be the dominant grain at about 4 \Rs \citep{Mukai1974} given the dust evaporation (i.e., dust-free zone), which is expected to be variable depending on the grain composition (and thus should be elongation dependent). 
\par

The reddest data point, 788.4 nm, is noticeably higher than the best fit power-law, perhaps indicating that the slope of the F-corona color spectrum may steepen in the infrared. Such a steepening is expected at longer wavelengths ($> 1 \mu$m) due to thermal emission of the dust (e.g., \citealt{Kimura1998}). The precise nature of the F-corona color at different elongations is not well known, though previous work has indicated a shallower slope in the Zodiacal light at high elongations compared to the F-corona (see \citealt{Koutchmy1985} and therein). It is difficult to make definitive statements on the nature of the F-corona from the existing observations, as most studies have only probed an inner elongation of about 4 \Rs. Thus, the nature of the inner F-corona is still largely unknown. Perhaps the color spectrum starts with a steeper slope in the inner corona and becomes shallower at higher elongations, which could be caused by thermal emission of particularly hot dust near the Sun at the edge of the dust-free zone. It is entirely possible that a mix of spatially variable grain species distributions and some component of thermal emission near $1 \mu$m may contribute to the precise color slope of the F corona in the near-infrared. Future work should endeavor to resolve the color spectrum of the F-corona across a wide range of wavelengths, and look for spatially dependent changes that would indicate variability in the thermal emission, mineralogy and/or size distribution of dust in the F-corona and Zodiacal light.
\par

An important consequence of the value of the color slope in this study, and in previous work, is that one must be careful when considering F-corona emission over different wavelengths. A polarization inversion method at 800 nm will certainly arrive at a different F-corona intensity than an identical one performed at 500 nm, by a factor of about $50 \%$. Such a color dynamic could also impact the inferred polarization ratio of the K-corona, as there will be a different background unpolarized brightness (due to the F-corona) at different wavelengths across the visible spectrum. One must account for both the color of the F-corona and K-corona (from limb-darkening, see Section \ref{LimbDark}) in order to correctly quantify K- and F-corona emission. 

\par

It is worth noting, that the $tB$ inferred by our color based inversion method is rather insensitive to the precise result of the F-corona color slope and F-corona calibration offset (see Section \ref{method}), since it is taken simply as the colorless component of emission between our observables across the visible spectrum (after accounting for limb-darkening effects, see Section \ref{LimbDark}). This colorless component remains almost exactly the same regardless of assumptions about the F-corona, which is a strength for this method of inversion. Any uncertainty in the calibration offset simply shifts the relative brightness of the F-corona (as addressed in Section \ref{iteration}), which acts to shift the entire image up or down by a constant value. Thus, the colorless component of the emission remains unchanged in the inference, while the colored component can be stretched based on the offset. Our final results are thus robust in inferring the K-corona $tB$ signal, while the inferred F/K offset only needs to assume axis-symmetric F-corona emission to converge quickly on the best fit offset. 

\newpage
\subsection{K-corona Total and Polarized Brightness}
\label{KcorBright}

\begin{figure*}[t!]
\centering
\epsscale{0.7}
\includegraphics[width = 5.5in]{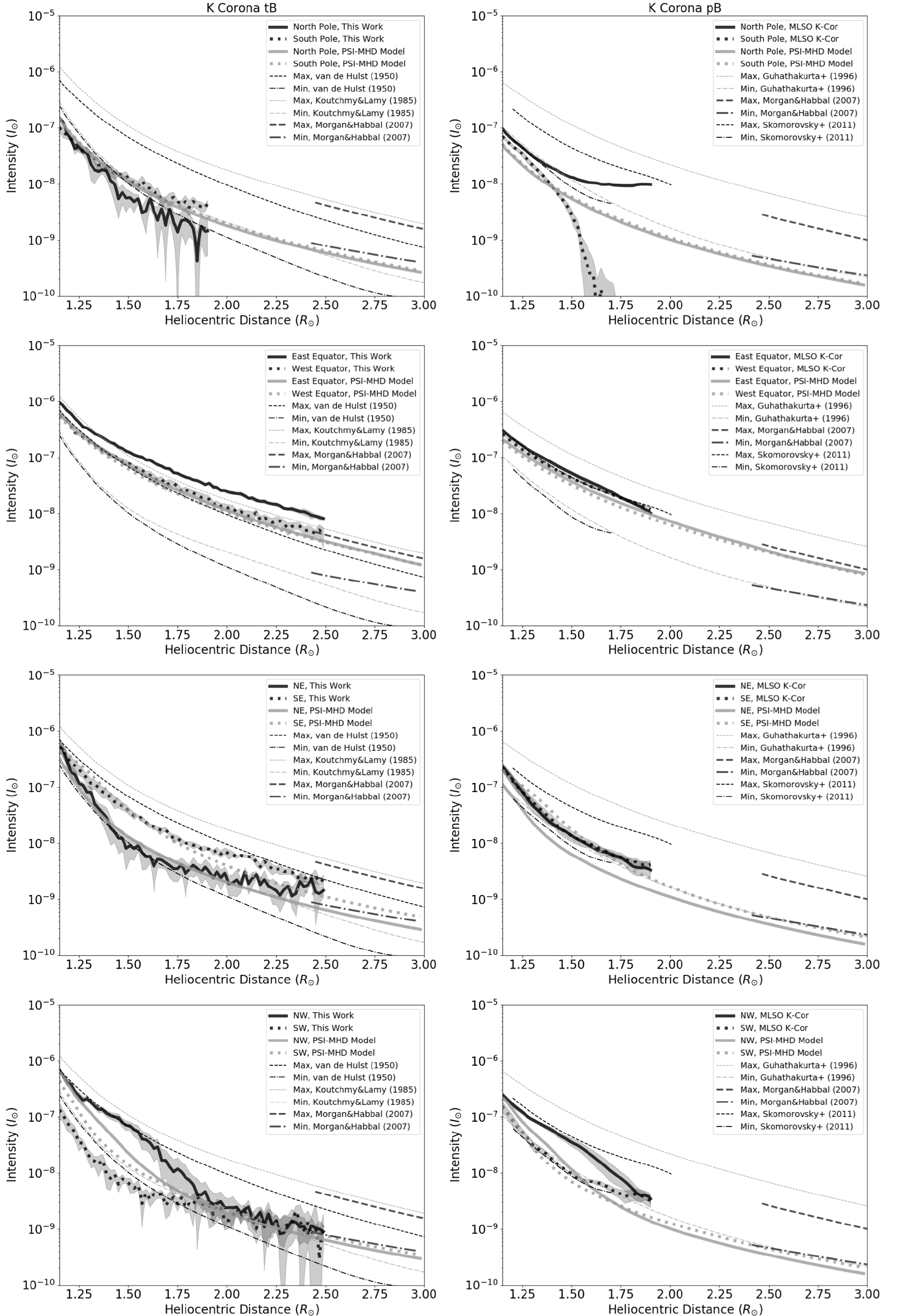}
\caption{Left panels show radial scans of the K-corona total brightness ($tB$) from the inversion technique (see Section \ref{method}), for a collection of different latitude regions. All scans are done inside $15^\circ$ wedges centered on the cardinal direction noted in the legends of each panel. The right panels show the same scans for the K-corona polarized brightness ($pB$) from MLSO's K-Cor (see Section \ref{Kcor}). A collection of literature values and the PSI-MHD model prediction (see Section \ref{PSI}) for this eclipse are also shown for comparison.} \label{FigKprofile}
\end{figure*}

\begin{figure*}[t!]
\centering
\epsscale{0.7}
\includegraphics[width = 5.5in]{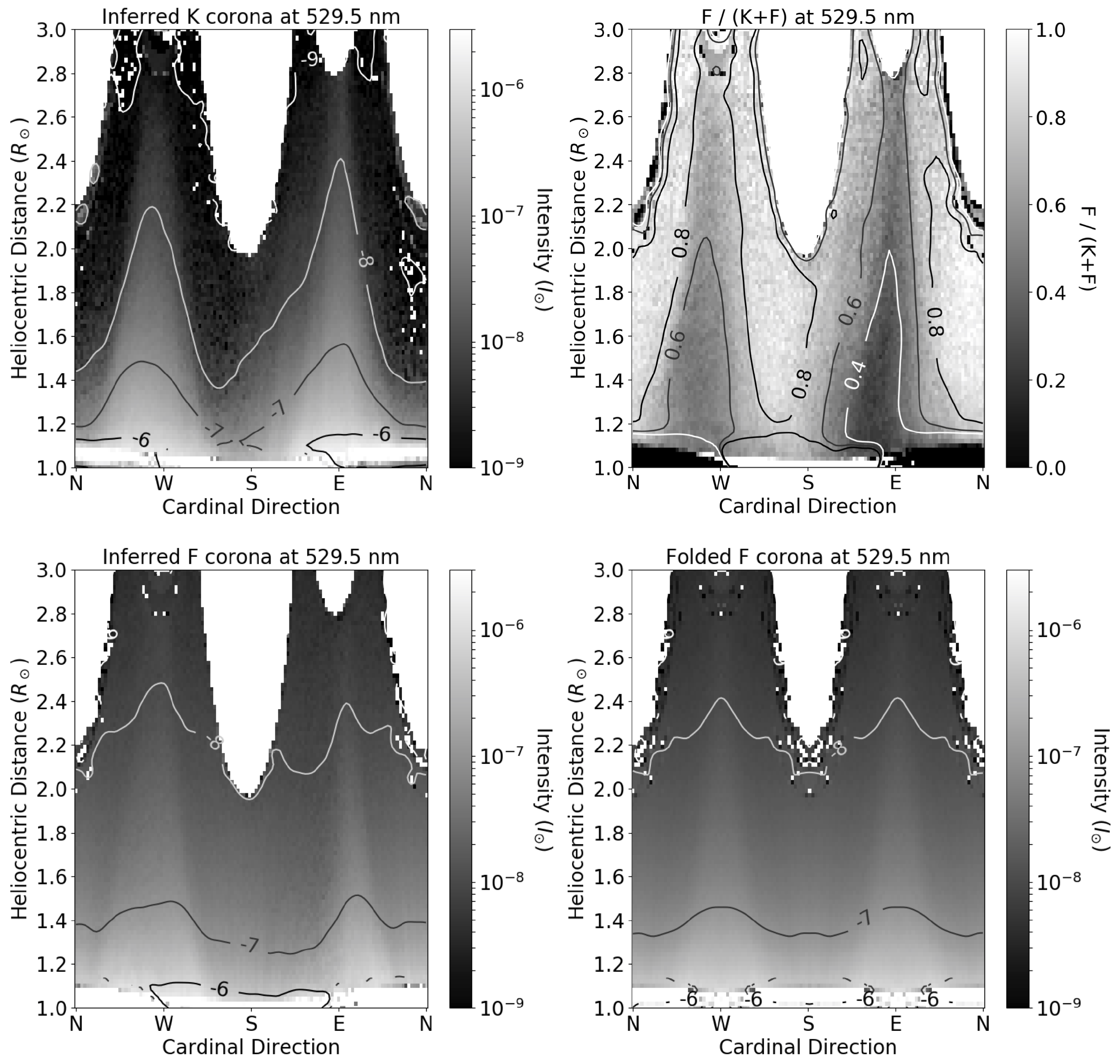}
\caption{Top left: Final inferred K-corona Total Brightness ($tB$), from the procedure described in Section \ref{method}. Top right: Inferred F/(K+F) ratio at 529.5 nm. Bottom left: Final inferred F-corona brightness at 529.5 nm. Bottom right: Axis-symmetric average of the F-corona inference.}
\label{FigF+K}
\end{figure*}

\begin{figure*}[t!]
\centering
\epsscale{0.7}
\includegraphics[width = 5.5in]{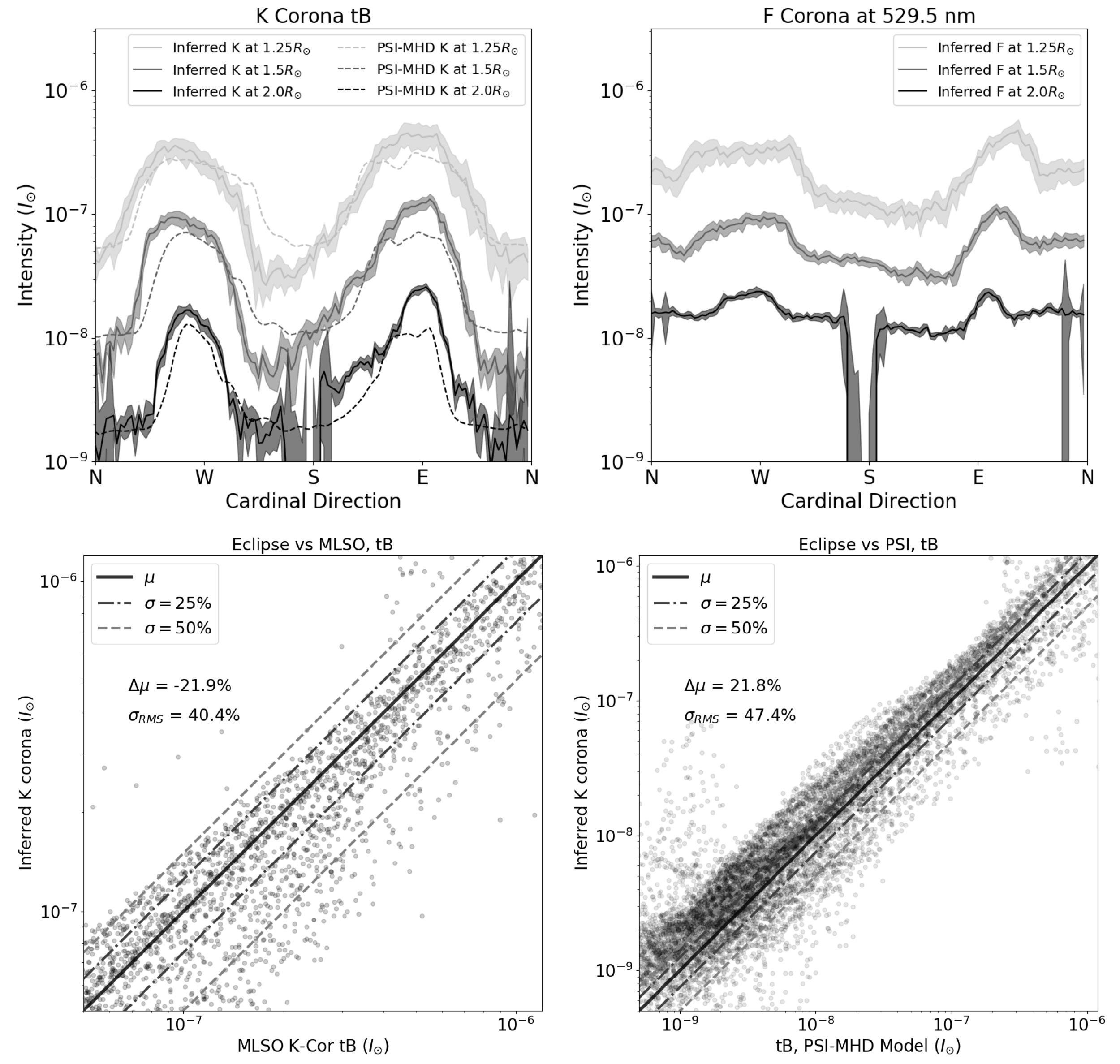}
\caption{Top left: Traces at 1.25, 1.5, and 2.0 \Rs \ (scan width of 0.1 \Rs) of the inferred K-corona $tB$ along with the PSI-MHD model prediction of $tB$ (see Section \ref{PSI}). Top right: Same as the top left, but for the inferred F-corona emission at 529.5 nm. Bottom left: Direct comparison between the inferred K-corona $tB$ and the K-Cor corrected $tB$ (from $pB$ and $pB$/$tB$, see Section \ref{Kcor}). Bottom right: Direct comparison between the inferred K-corona $tB$ and the PSI-MHD model prediction of $tB$.}
\label{FigCompare}
\end{figure*}

The K-corona total brightness ($tB$) inferred from the 2019 July 2 total solar eclipse (see Section \ref{Eclipse}) is shown in a collection of radial scans in the left panels of Figure \ref{FigKprofile}, along with some comparisons to the literature and the PSI-MHD model (see Section \ref{PSI}). The spatially resolved K-corona inference is given in the top left panel of Figure \ref{FigF+K}. Traces of the K-corona intensity at a set of fixed radial distances are shown in Figure \ref{FigCompare}, in addition to direct LOS intensity comparisons between the eclipse, the MLSO K-Cor observations and the PSI predictions. 
\par

In general, we find a K-corona $tB$ that is consistent with the range of values typically found in the corona. For example, the streamers have a substantially higher ($\sim 10 \times$ at 1.5 \Rs) intensity than the poles, as expected, and the highest and lowest intensities broadly agree with the higher and lower estimates from the literature (i.e., \citealt{vandeHulst1950, Koutchmy1985, Morgan2007}). Similarly, the MLSO K-Cor observations (see Section \ref{Kcor}) on the day of the eclipse are consistent with previous work on the polarized brightness ($pB$) of the K-corona (i.e., \citealt{Guhathakurta1996, Morgan2007, Skomorovsky2011}), which are shown in the right panels of Figure \ref{FigKprofile}. The comparison to \cite{Skomorovsky2011} is especially interesting, as they observed the K-corona $pB$ during the 2008 total solar eclipse, close to solar minimum (much like this 2019 eclipse). Further, the MLSO K-Cor data appear to be rather robust in the equatorial streamers out to about 1.8 \Rs, but they diverge rather quickly in the lower intensity regions near the poles of the Sun (top right panel of Fig. \ref{FigKprofile}), and perhaps are only accurate to a distance of about 1.2--1.3 \Rs. Regardless, the MLSO dataset is a valuable and essential component to this work (as the calibration source and initial guess of the K-corona), and can be used for detailed analysis of the lower regions of the corona. 

\par
Of course, there are a multitude of studies that have observed the K-corona $pB$ and $tB$ intensity, either through direct observation of the F-corona lines (e.g., \citealt{vandeHulst1950}) or via polarization observations. The polarization observations, in particular, have been repeated many times with both eclipses (e.g., \citealt{Guhathakurta1996, Skomorovsky2011}) and coronagraphic datasets (e.g., \citealt{Morgan2007, Lamy2020}). To compare our K-corona intensity results with other studies, we have chosen a small number for the sake of brevity and clarity, especially given that the decades of work have consistently reproduced the same intensities within a range that is comparable to the real physical variation of the electron density over the solar cycle and across different coronal structures (e.g., streamers vs. coronal holes).

\par 

In addition to finding a K-corona intensity that is consistent with previously published values, we find that the PSI model for this specific eclipse matches the data reasonably well. A direct comparison for each LOS between our inferred K-corona $tB$ and both the $pB$/$tB$ corrected K-Cor data (as in Figure \ref{figPSI}) and the PSI-MHD model prediction are shown in the bottom panels of Figure \ref{FigCompare}. We find a very close match between the PSI model and our inferred K-corona $tB$, with only an $21.8 \%$ average disagreement down to a brightness of $10^{-8} \ I_\odot$. The scatter between the datasets is substantially higher, at about $47.4 \%$. This scatter is expected even if the physics of the model were perfect, because the coronal morphology of the model is largely determined by the surface boundary conditions which are comprised of observations made in the weeks prior to the eclipse (as mentioned in Section \ref{PSI}). In this sense, the surface magnetic field of the model cannot perfectly match the instantaneous surface flux distribution during the eclipse, since some discrepancies between streamer shapes and locations are expected. The polar magnetic flux, which has strong effects on the precise angle of streamers in the final prediction \citep{Riley2019}, is also not well constrained. This collection of effects is more than sufficient to explain the scatter between the model and our data. 

\par
There is an interesting anomaly between our inferred K-corona $tB$ and the PSI model prediction in the North-West scan (bottom left panel of Fig. \ref{FigKprofile}), where our inferred K-corona remains at a higher intensity until a distance of about 1.8 \Rs. This `bump' in intensity might seem as though there could be some issue with the observational data, but we see a bump in the same location in the K-Cor $pB$ observation (bottom right panel of Fig. \ref{FigKprofile}). Seeing this feature in both of the entirely independent observations hints that this is some real feature in the corona that is not seen in the model, indicating a morphology or boundary condition mismatch.

\par

Once the K-Cor data are corrected with the $pB$/$tB$ fraction from the PSI model (see Section \ref{PSI}), they match the eclipse data very well. There is an average difference of $21.9\%$ between the eclipse and K-cor datasets, with the inferred eclipse intensity being fainter than K-cor. This difference could be explained by an additional $pB$ signal that is contaminating the K-Cor data, such as some unaccounted $pB$ signal from the Earth's atmosphere, or could perhaps indicate a slight $pB$ signal from the F-corona for low elongations (see Section \ref{FcorBright}). It is also worth considering that the overall scatter in the cross-match is $40.4 \%$, which could imply that photometric uncertainties and/or issues in the model $pB$/$tB$ correction may be sufficient to explain the difference between these datasets. In the next section, we expand on this possibility that the F-corona itself may have a slight $pB$ component in the lower regions of the corona which is causing this $pB$ excess, as part of a broader F-corona intensity discussion.

\subsection{F-corona Brightness}
\label{FcorBright}

Our inference of the F-corona intensity at 529.5 nm, is shown in Figure \ref{FigF+K}. The bottom left panel of the figure shows the spatially resolved F-corona map, while the bottom right panel shows the same data, except with folding across the polar and equator axis. The folded F-corona looks rather similar to the unfolded one, so the folded version is likely a good estimate of what the F-corona really looks like. This similarity supports the initial assumption for using the folded F-corona as a method for inferring the F-corona offset in the calibration procedure (see Section \ref{iteration}). We find that the F-corona is slightly higher near the equator than at the poles, as demonstrated by the F-corona intensity traces shown in the top right panel of Figure \ref{FigCompare}. This difference between the polar and equatorial F-corona has been noted before (e.g., \citealt{Koutchmy1985,Morgan2007,Lamy2020}), though older work including \cite{vandeHulst1950} and others of that period assumed that the F-corona was uniform for all solar latitudes, which is increasingly clear to be incorrect.  

\par
There are some slight anisotropies between the Northern and Southern polar regions, where the Northern latitudes seem to have a higher F-corona intensity and a lower K-corona intensity (shown in Figure \ref{FigF+K}, see Section \ref{KcorBright}). It is perhaps possible that there could be some variation in the inner regions of the F-corona due to a physical interaction between the K- and F-corona (i.e., collisions, B field, etc.). It could also be interpreted as implications on the dust-size or grain-composition distributions or variation in the dust-free zone. However, it is more likely that this is an effect from slight errors in the synthesis procedure of this work, as the variance between similar latitudinal regions is rather small compared to the overall differences between the F- and K-corona profiles at different distances. More data of this type are required across a solar cycle to investigate if there are small changes in the F-corona that correlate with K-corona emission.
\par

The relative F-corona intensity of the total continuum signal is shown in the top right panel of Figure \ref{FigF+K}. These F/(K+F) values are quite high, at about 0.3 to 0.5 at the lowest in the eastern streamer, then rising to about 0.8 at a distance as low as 1.25 \Rs \ in the polar coronal holes. This large contribution of the F-corona partly explains why the polarization observations of K-Cor (see Sections \ref{Kcor} and \ref{KcorBright}) are not able to probe farther than about 1.3 \Rs \ in the polar regions, as the vast majority of the photometric signal is originating from the unpolarized (or perhaps slightly polarized) F-corona. The high F-corona fraction is no doubt a result of the Sun being almost precisely at solar minimum during the 2019 July 2 total solar eclipse. We expect the electron density to be at its lowest overall in the corona during this time, while the F-corona has been observed to be rather invariant over the solar cycle \citep{Morgan2007}. The high F-corona fraction during this solar minimum corona also increases the confidence in our ability to quantify the color of the F-corona between the channels (as presented in Section \ref{FcorColor}), as it is a large fraction of the overall intensity. The results of this work could thus be generalized to other periods of the solar cycle to remove the F-corona signal and enable a more complete study of the total K-corona intensity.

\par

\begin{figure*}[t!]
\centering
\epsscale{0.7}
\includegraphics[width = 5.5in]{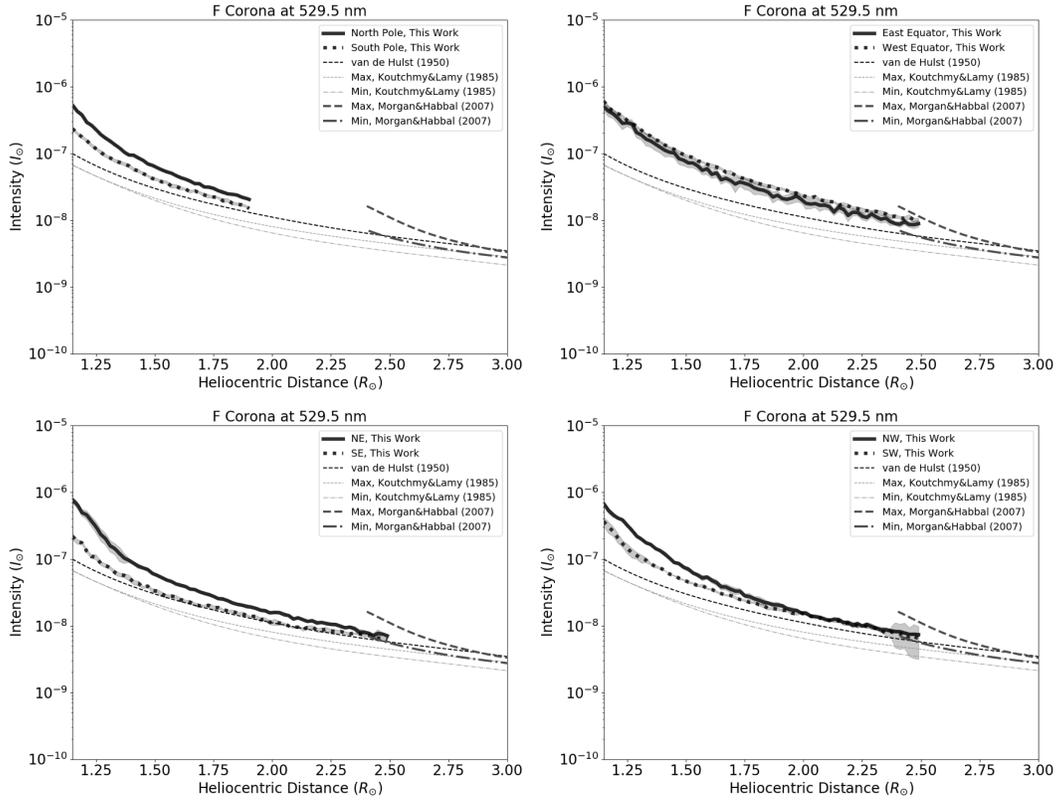}
\caption{Radial scans of the F-corona total brightness from the inversion technique (see Section \ref{method}), for a collection of different latitude regions (same style as Fig. \ref{FigKprofile}). All scans are done inside $15^\circ$ wedges centered on the cardinal direction noted in the legends of each panel. A collection of literature values are also shown for comparison.}
\label{FigFprofile}
\end{figure*}

The overall intensity of the F-corona agrees reasonably well with previous published work, as can be seen in the radial scans shown in Figure \ref{FigFprofile}. We do find that the F-corona is considerably brighter than \cite{Koutchmy1985}, but is rather similar to \cite{vandeHulst1950} in the coronal poles. Our results match rather well with \cite{Morgan2007}, who inferred the F-corona via $pB$ and $tB$ observations of LASCO-C2 (thus at high elongations, $> 2.4$ \Rs). The historical work of \cite{vandeHulst1950} used Fraunhofer line depths (see Section \ref{intro}) in contrast to the polarization method of \cite{Koutchmy1985} and \cite{Morgan2007}. At high elongations, it has been observed that the F-corona is essentially unpolarized (e.g., see \citealt{Koutchmy1985,Lamy2020} and therein), but it is somewhat uncertain in the low corona. However, recent work by \cite{Morgan2020} did infer a slight polarization fraction in the F-corona of around 5--8$\%$ at a heliocentric distance of 4 to 5.5 \Rs, using LASCO-C2 observations.

\par

If the F-corona does have a substantial polarization fraction in its emission at low elongations, there could be a bias in the $pB$ inferred F-corona intensity and thus would explain why our color-$tB$ inversion method is finding a higher F-corona intensity, but would not impact the F-line depth analysis of \cite{vandeHulst1950}. The lower regions of the corona also have the lowest polarization fraction from the K-corona (see Figure \ref{figPSI}), so any contribution from a small polarized dust signal in the F-corona at low elongations could contribute excess intensity to $pB$ observations that is normally attributed entirely to the K-corona. If the F-corona is indeed dominated by diffracted light from large particles (compared to visible wavelengths), then \cite{Lamy2020} claims it is safe to assume essentially no polarized component of F-corona emission. If there are perhaps some smaller dust grains near the Sun that are directly scattering (rather than only a diffraction contribution), then we would expect some amount of F-corona polarization near the Sun (per the argument of \citealt{Koutchmy1985}). 
\par
The excess of our inferred F-corona signal for the low corona (based on the color), combined with the excess of $pB$ intensity seen in the MLSO K-Cor data compared to the eclipse data (see Section \ref{KcorBright}) both seem to indicate that the F-corona may be slightly polarized. The MLSO K-Cor data are also taken between 720 to 750 nm, where we have demonstrated that the F-corona is a reasonably large fraction of the total brightness; this eclipse being almost exactly at solar minimum enhances any contribution from the F-corona due to the low electron density of the corona at solar minimum. 
\par

There could also be some solar cycle variation of the F-corona in the very low corona. Some sort of physical (i.e., $n_e, T_e, B,$ etc.) variation in the corona driven by the solar cycle could possibly shift the distance of the dust-free zone (see Sections \ref{intro}, \ref{FcorColor}). If the dust-free zone moves inward during solar minimum, we would expect more scattered light in the corona near the Sun (even if it is still somewhat out of the plane of sky) -- which would also enhance any $pB$ effects that may be present in the F-corona, as well as increase the overall F-corona intensity. Additional studies are no doubt required to test these possibilities further, especially one where both the color and $pB$ brightness can be observed simultaneously at a total solar eclipse. Off-limb observations by the new Daniel K. Inouye Solar Telescope (DKIST; \citealt{Tritschler2016}) observatory may also help to address this problem.

\section{Conclusions}
\label{conc}

We have presented a new inversion technique (Section \ref{method}) which extracts the K- and F-corona total brightness ($t_B$, see Sections \ref{KcorBright} and \ref{FcorBright}) and F-corona color (see Section \ref{FcorColor}) using a set of narrowband total solar eclipse (TSE) observations across the visible spectrum (see Section \ref{Eclipse}). Our results are broadly consistent with previous work, as well as the MLSO COSMO K-Cor coronagraph observations (see Section \ref{Kcor}) and the state-of-the-art PSI-MHD model prediction (see Section \ref{PSI}). We find that the eclipse and K-Cor data are roughly in agreement out to $\sim 1.3$ \Rs \ in the coronal holes, and up to 1.8 \Rs \ in the streamers, highlighting that the MLSO COSMO K-Cor data are a robust dataset for studying the lower regions of the corona, in the absence of an eclipse.
\par
The major conclusions from this work include:
\begin{enumerate}
\item The F-corona color spectrum is found to be consistent with that of \cite{Koutchmy1985}, with a best-fit power law of $I_F \propto \lambda^{0.91 \pm 0.07}$ for visible wavelengths. 
\item The F-corona brightness is found to be non-uniform with latitude (as noted in previous studies), where the equatorial regions have a higher F-corona intensity compared to the poles. 
\item We find that the F-corona brightness is significantly higher in the low corona (see Fig. \ref{FigFprofile}) than found in previous studies; but our inference matches the literature very well beyond about 1.5 \Rs. This finding may indicate that the nature of the dust-free zone varies with solar cycle, or perhaps that the F-corona has a polarized brightness component that is unaccounted for in $pB$ driven inversions.
\item The eclipse driven K-corona inversion matches the PSI-MHD model very well out to high heliocentric distances. The close match between the model and observations indicates that the model is doing an excellent job at predicting the K-corona, and thus is likely generating a reasonable electron density for the corona in 3D. 
\end{enumerate}
\par
The results presented here support the need for further photometric color analysis of the corona as a tool to study the F-corona. There are potentially unexplored avenues to constrain the dust-free zone, as well as the dust distribution and the far-versus-near component of the F-corona scattering (i.e., scattering vs. diffraction, see Section \ref{FcorColor}). Further, this study emphasizes the importance of narrowband observations of the corona which could be performed with future ground- and space-based coronagraphs. These coronagraphs should utilize narrowband filters to probe not only ionic line emission, but also the color of the coronal continuum emission. Even with recent advancements in instrumentation, TSEs continue to provide unparalleled access to the corona continuously from just above the solar surface out to at least 2.5 \Rs. At present, no existing telescope can rival the coronal view provided by a total solar eclipse.

\subsection*{Acknowledgments}
We thank all the observers that performed the narrowband observations at the 2019 July 2 total solar eclipse (see Table \ref{tableEclipse}). The K-Cor data were courtesy of the Mauna Loa Solar Observatory, operated by the High Altitude Observatory, as part of the National Center for Atmospheric Research (NCAR). NCAR is supported by the National Science Foundation.
\par
Financial support was provided to B. Boe by the National Science Foundation under Award No. 2028173, and by AURA/NSO with grant N97991C to the Institute for Astronomy at the University of Hawaii. S. Habbal and the 2019 eclipse expedition were supported under NSF grant AST-1733542 to the Institute for Astronomy of the University of Hawaii. C. Downs was supported by NASA grants 80NSSC18K1129 and 80NSSC20K1285. M. Druckm\"uller was supported by the Grant Agency of Brno University of Technology, project No. FSI-S-20-6187. 
\par

\bibliographystyle{apj}

\end{document}